\documentclass[11pt]{article}
\usepackage{amsmath,amssymb,amsthm,amsfonts,graphicx,graphics,epsfig,verbatim,array}

\usepackage{setspace}

\newcommand{\beq}{\begin{equation}}
\newcommand{\eeq}{\end{equation}}
\newcommand{\beqr}{\begin{eqnarray}}
\newcommand{\eeqr}{\end{eqnarray}}
\newcommand{\beqrn}{\begin{eqnarray*}}
\newcommand{\eeqrn}{\end{eqnarray*}}
\newcommand{\beqn}{\begin{equation*}}
\newcommand{\eeqn}{\end{equation*}}
\newcommand{\bei}{\begin{itemize}}
\newcommand{\eei}{\end{itemize}}
\newcommand{\bes}{\begin{small}}
\newcommand{\ees}{\end{small}}
\newcommand{\bec}{\begin{center}}
\newcommand{\eec}{\end{center}}

\newcommand{\al}{\alpha}

\newcommand{\tht}{\theta}

\newcommand{\bQ}{{\bf Q}}
\newcommand{\bbf}{{\bf f}}
\newcommand{\IFF}{{I_F^{\text{mean}}}}
\newcommand{\IFV}{{I_F^{\text{var}}}}
\newcommand{\IFr}{{I_F^{\text{corr}}}}
\newcommand{\rij}{\rho_{i,j}}
\newcommand{\IFcov}{I_F^{\text{cov}}}

\newcommand{\rtwo}{\rho_{i,j}}
\newcommand{\defn}{\overset{ \text{def} }{=}}

\title{Stimulus-dependent correlations and population codes}

\author{
Kre\v{s}imir Josi\'{c}\\
Department of Mathematics\\
University of Houston \\ Houston TX 77204-3008, USA
\and 
 Eric
Shea-Brown \\
Department of Applied Mathematics \\
University of Washington  \\
Seattle, WA 98195-2420 \and
Brent Doiron \\
Department of Mathematics \\ 
University of Pittsburgh \\
Pittsburgh, PA, 15206 \and
Jaime de la Rocha\\ 
 Center for Neural Science 
 \\ New York University \\ New York NY 10012, USA
 }

\begin{document}

\singlespacing

\maketitle

\begin{abstract}
The magnitude of correlations between stimulus-driven responses of pairs of neurons can itself be stimulus-dependent.   We examine how this dependence  impacts the information carried by neural populations about the stimuli that drive them.  
Stimulus-dependent changes in correlations can both carry information
directly and modulate the information separately carried by the firing rates and variances.
We use Fisher information to quantify these effects and show that, 
although stimulus dependent correlations often carry little information directly, their modulatory effects on the overall information can be large.
In particular, if the stimulus-dependence is such that correlations increase with stimulus-induced firing rates, this can significantly enhance the information of the population when the structure of correlations is determined solely by the stimulus.  However, in the presence of  additional strong spatial decay of correlations, such stimulus-dependence may have a negative impact.  Opposite  relationships hold when correlations decrease with firing rates.
\end{abstract}

\section{Introduction}

The impact of correlations on information
encoded in neural tissue is a subject with a substantial history.
We start our discussion with~\cite{zohary94},
which reported significant correlations between
neuronal responses in paired recordings of neurons in the visual area of monkeys.
Correlations were deemed undesirable, as they  lead
to a  \emph{decrease} in the signal-to-noise ratio of the
summed population activity~\cite{johnson80,britten92}.
Despite this impact on the signal-to-noise ratio, correlations in the neural response can \emph{increase}
the information that a population of neurons carries about a
signal~\cite{abbott99}.     The impact of correlations on coding depends in a complex way on their distribution over the neuronal population~\cite{romo03,chen06,roelfsema08,oram98,averbeck06,Ser+04,kohnprep,abbott99,shamir04,shamir06,sompolinsky01}.  As the range of potential patterns of correlation  is vast, and has not been characterized in most neurobiological systems, 
the effect of correlations is not fully understood.  


 In many studies to date, the correlation coefficient between
the responses of pairs of neurons was assumed to be independent of the stimulus driving the response.  In particular, it was assumed that covariances between cell responses change together
with the variance so that the correlation coefficient remained constant.  Information about stimulus identity could then be encoded solely in the rate and variability  of single cell responses~\cite{abbott99,shamir04,shamir06,sompolinsky01}.  However, experimental findings suggest that correlations themselves vary with stimuli \cite{deC+96, samonds03,kohn05,rocha06,Gra+89,rocha06, biederlack07,chacron08}.  More specifically, it has been shown in~\cite{kohn05} that correlations in the visual cortex (V1) vary with the stimulus orientation and contrast.  In~\cite{biederlack07}, it was demonstrated experimentally that in certain situations changes in perceived
brightness are related to changes in neural correlations.
Responses to prey-like vs.
conspecific-like stimuli  in electric fish have also been demonstrated to evoke
responses with different correlation structure~\cite{chacron08}. 


Here, we concentrate on a particular form of stimulus-dependence, in which correlations depend on stimulus-evoked firing rates (although many of our formulas hold more generally).  In recent work, we have shown that spike-to-spike correlations due to common inputs increase with firing rate for neural models and \emph{in vitro} neurons~\cite{rocha06}.  This effect was observed \emph{in vivo} in the anesthetized visual cortex~\cite{kohn05,greenberg08} and, in certain experimental regimes, for motoneurons \emph{in vitro}~\cite{Bin+01}.  In the oculomotor neural integrator the opposite effect was observed: correlations decreased with rate~\cite{AksayBST03}, perhaps due to recurrent network interactions.  We will study both of these cases, illustrating strongly differing effects of stimulus-dependence in each.

The goal of this paper is to examine, from a theoretical perspective, the impact of 
stimulus-dependent correlations on population coding.  Previously,  changes in discriminability due to changes in the covariance matrix of pairs of cells and small (3-8 cell) ensembles were examined by~\cite{AverbeckL03}.  Also, a series expansion of mutual information  to isolate and quantify the effects of stimulus-dependent correlations has been developed~\cite{PanzeriSTR99}. Similarly,~\cite{kohn07} use mutual information to assess the impact of tuned correlations measured in primate V1.  We take a somewhat different approach based on computing the impact of the stimulus dependence of correlations on the Fisher information ($I_F$) for populations of neurons whose response is described by tuning curves~\cite{abbott99}.

There are at least two distinct ways in which the stimulus dependence of correlations can impact
Fisher information.  
First, the fact that patterns of correlation across a population are adjusted as stimuli change can have a strong ``modulatory" impact on the information that {\it other} features of the neural response -- such as firing rates -- carry about the stimulus~\cite{kohn07,gutnisky08}.
We refer to this effect as {\it correlation shaping}.  To better understand this, note that a stimulus-independent correlation structure may be optimized for one stimulus.  However,  stimulus-dependence offers the possibility that the correlation structure is adjusted, and optimized, for a range of stimuli~\cite{averbeck06}.  In a related effect, adaptation has been shown to modify correlation structure and increase $I_F$~\cite{kohnprep,gutnisky08}.

Secondly, information may be encoded directly by changes in  the level of
correlation between neurons, in addition to encoding via changes in firing rate and variance.  We refer to this mechanism as {\it correlation coding.}  One scenario where correlation coding clearly dominates if stimuli {\it only} affect the correlation structure, leaving rates and variances relatively constant, as has been observed experimentally~\cite{vaadia95,biederlack07,chacron08}. 

The balance of the paper proceeds as follows. We start by defining our statistical description of the neural response to stimuli in Section~\ref{S:fisher}.  The information in the response
of two cells is studied in Section~\ref{s.twocells}.  As we show, the insights gained from this case
can be extended to small populations, but do not always apply to larger populations.
In Section~\ref{S:arbitrary} we study the information in the response of a large population.
Here, we find that correlation shaping effects can be substantial, and often dominate over correlation coding. 
In Section~\ref{s.decay} we extend the model to address additional structure of correlations across the population, by including  decay of correlations that depends explicitly on the spatial or ``functional" distance between preferred stimuli of neurons, as  shown experimentally.   We find that  the impact of correlation shaping in the presence of such a decay continues to be strong, but that correlation coding also plays a significant role.  We conclude with a discussion of the results.  A number of analytical results used in the main body of the paper, which may be of independent interest, are derived in the appendices.

\section{Setup} \label{S:fisher} 

\paragraph{Structure of correlations}
We consider a population of $N$ neurons responding to a stimulus described by a scalar variable $\theta$ (for example, the orientation of a visual grating).
The number of spikes  fired
by neuron $i$ in response to  stimulus $\theta$ during a fixed  time interval 
is given by
\begin{equation}\label{E:rate}
r_i(\theta) = f_i(\theta) + \eta_i(\theta),
\end{equation}
where $f_i(\theta)$ is the mean  response of neuron $i$ across trials, and
$\eta_i(\theta)$ models the trial--to--trial variability of the response.  
We use boldface notation
for vectors, so that $\boldsymbol{r}(\theta)$ denotes the multivariate
random variable  $\boldsymbol{r}(\theta) = [r_1(\theta), r_2(\theta), \ldots, r_N(\theta)]^T$. For simplicity, we sometimes
suppress  dependences on $\theta$.  

We assume that $\boldsymbol{\eta}$ follows a
multivariate distribution with zero mean and covariance matrix
$\bQ(\theta)$ defined by
\begin{equation} \label{E:crosscor}
Q_{i,j}(\theta) = \delta_{i,j} v_i(\theta) + ( 1- \delta_{i,j}) \rij(\theta) \sqrt{v_i(\theta) v_j(\theta)}.
\end{equation}
Here $v_i(\theta)$
is the variance of the response of  cell $i$, and $-1 \leq \rij(\theta) \leq 1$ is the correlation coefficient of the response of
cells $i$ and $j$.  Although most of our results will be discussed in the range of small to intermediate
correlations, $\rij \lesssim 0.5$, a similar analysis can be used to study the behavior of populations
close to perfect correlations, $\rho_{i,j}  \approx 1$.  Assumptions on the form of the distribution, beyond this covariance, are made only where needed.

For studies of stimulus-dependent correlations in small-to-intermediate populations (Section~\ref{s.twocells}), we will allow for general forms of $\rij(\theta).$  When we study large populations (Sections~\ref{S:arbitrary} and \ref{s.decay}),  we will assume that   
\begin{equation} \label{E:rij}
\rij(\theta) = S_{i,j}(\theta) \; c(\phi_i - \phi_j),
\end{equation}
where $\phi_i$ and $\phi_j$ are the preferred stimuli of neurons $i$ and $j$ respectively.  
The stimulus independent term $c(\phi_i - \phi_j)$ represents the \emph{spatial or functional} structure of correlations
in the population.    It describes how correlations vary across the population according to their preferred stimuli, perhaps due to hardwired differences in the level of shared inputs.    For instance, neurons which prefer similar stimuli
are frequently closeby in the cortex, and may share a larger number of common inputs
than neurons that exhibit different preferences~\cite{zohary94,lee98}.  Moreover, the set of neurons upstream of two cells with similar stimulus preferences may also undergo common fluctuations in their activity.   Therefore, $c(\phi_i - \phi_j) = c(\Delta \phi)$ is frequently assumed to decrease with the functional distance $\Delta \phi$. We will refer to this simply as ``spatial decay"~\cite{dayan2001,sompolinsky01,wilke02}.

We emphasize that it is the stimulus dependence of the correlation coefficient, $\rij(\theta)$, that distinguishes the present work from several previous investigations~\cite{abbott99,sompolinsky01,shamir04}.  This dependence enters through the term $S_{i,j}(\theta)$~\cite{kohn05,rocha06,greenberg08}.  We mainly investigate cases in which correlations between pairs of cells increase, decrease, or have a single maximum with respect to the evoked firing rates $f_i$ and $f_j$~\cite{rocha06,brown07,kohn05,Bin+01,AksayBST03}. However, our results could also be applied to cases with other relations between $\rho_{ij}(\theta)$, $f_i$, $f_j$, $v_i$, and $v_j$ such as those arising for different circuit and nonlinear spike generation mechanisms (cf.~Fig. 4 of~\cite{rocha06}).  

For large populations, we extend the multiplicative model in~\cite{shamir01} 
to the case of stimulus-dependent correlations by assuming that
	\beq
	S_{i,j}(\theta) = s_i(\theta) s_j(\theta) \;,
	\label{e.rhoform}
	\eeq
	where $-1 < s_i(\theta),s_j(\theta) < 1$.
Here $s_i(\theta)$ may be thought of as the propensity of a neuron's response to be correlated, and $s_i^2(\theta)$ as the correlation between two neurons which  respond equivalently to the stimulus.  There are several reasons for adopting the form given in Eq.~\eqref{e.rhoform}.  Firstly, this form of $\rho_{ij}$ arises  for small to intermediate correlation in neuron models producing a spike train with renewal statistics~\cite{rocha06,brown07}.   Moreover, in this case correlation has also been shown to vary with the geometric mean of the firing rate of pairs of cells {\it in vivo} \cite{kohn05,rocha06}, which can be modeled using Eq.~\eqref{e.rhoform}.  Additionally, this form keeps the computations at hand analytically tractable for large population sizes, and  limits the number of cases under study.

\paragraph{Fisher information}  To quantify the fidelity with which a neuronal population represents a signal,
we use Fisher information~\cite{seung93,dayan2001}.  For the probability distribution 
$p[{\bf r} | \theta]$  of the spike count vector
$\bf{r}$  given stimulus $\theta$,  the Fisher
information is defined as
$$
I_F(\theta) = \left<- \frac{d^2}{d \theta^2} \log  p[{\bf r} | \theta] \right>,
$$
where $< \cdot >$ denotes expectation over the responses ${\bf r}$.
The inverse of the Fisher information, $1/I_F(\theta)$, provides a lower bound on the variance (i.e., an upper bound on the accuracy) of an unbiased decoding estimate of $\theta$ from the population response~\cite{cover91,dayan2001}.  Fisher information is directly related to the discriminability $d'$ between
two stimuli $\theta$ and $\theta+\Delta \theta$, since $d' \approx \Delta \theta \sqrt{I_F(\theta)}$
for small $\Delta \theta$~\cite{dayan2001}.  

The Fisher information
can be written as~\cite{kay93}
\beq
I_F = \IFF + \IFcov.
\eeq
Here
\beq 
\IFF = \bbf'^T \bQ^{-1} \bbf'  \label{E:IFFgeneral}
\eeq
is known as the ``linear approximation" to the Fisher information, or  the linear Fisher information.  Specifically, the inverse of $ \bbf'^T \bQ^{-1} \bbf' $ gives the asymptotic\footnote{Here, asymptotic implies that the optimal linear estimator is constructed based on full knowledge of the mean and covariance of the underlying stimulus-response distributions.} error of the optimal linear estimator of the stimulus, for a response to the stimulus that follows {\it any} response distribution that has mean $\bbf(\theta)$ and covariance $\bQ(\theta)$~\cite{Rao45,Cramer46,Ser+04}.  In particular, this applies to gaussian or nongaussian distributions.  

The second term, $\IFcov$, does depend on the the form of the response distribution, beyond its covariance.  In the following, whenever computing $\IFcov$, we assume that $\boldsymbol{\eta}$ follows a multivariate Gaussian distribution, so that~\cite{kay93}
\begin{equation} \label{E:fullfisher}
\IFcov= \frac12\text{Tr}\left[ \bQ' \bQ^{-1}\bQ' \bQ^{-1} \right].
\end{equation}
As we explain below, correlation coding affects only $\IFcov$, while correlation shaping affects 
both $\IFF$ and $\IFcov$.  

\section{The cases of cell pairs and small populations}
\label{s.twocells}

We start by considering the impact of correlations on the information carried
by cell pairs, and small populations ($N < 1/\rho$).  This was the setting of 
many experimental studies which addressed the role of correlations in the neural code~\cite{petersen01,averbeck03,rolls03,samonds04,roelfsema08,gutnisky08}.  We use analytical
expressions  to show that, depending on correlation structure,  correlation shaping can have either a positive or negative impact on $\IFF$.  
Most beneficial are high correlations between neurons with different
stimulus preferences, and low correlations between neurons with similar preferences.  For small to intermediate correlations, $\IFF \approx I_F$, and hence correlation coding has little effect.  These results are in agreement with previous observations~\cite{rolls03,averbeck04,averbeck06}. 
We emphasize that these results can be expected to hold only when $N<1/\rho$:  in subsequent sections we show that the
intuition gained from studying cell pairs may not always extend to larger populations.

\paragraph{Fisher information in cell pairs}  We first consider two cells whose
response follows a bivariate Gaussian distribution given by Eqs.~(\ref{E:rate}--\ref{E:crosscor})
For two neurons, we write the correlation coefficient as $\rho_{1,2} = \rho_{2,1} = \rho$, and
obtain
\begin{equation} \label{E:exact}
\begin{split}
I_F &= \underbrace{ \frac{1}{1 - \rho^2} \left[ \frac{f_1'}{\sqrt{v_1}} - \frac{f_2'}{\sqrt{v_2}} \right]^2   +\frac{2}{1+\rho} \left[ \frac{ f_1' f_2' }{\sqrt{v_1 v_2}} \right]}_{\IFF}  \\
&+ \underbrace{\frac{2 - \rho^2}{4(1 - \rho^2)} \left[ \left(\frac{v_1'}{v_1}\right)^2 + \left(\frac{v_2'}{v_2}\right)^2\right] - \frac{\rho^2}{2(1- \rho^2 )} \frac{v_1' v_2'}{v_1 v_2}
+ \left[ \frac{ (1 + \rho^2)\rho'}{1 - \rho^2}\right] \left[ \frac{\rho'}{1 - \rho^2} - \frac{ \rho}{1 + \rho^2}
\left(\frac{v_1'}{v_1} + \frac{v_2'}{v_2}\right) \right] }_{\IFcov},
 \end{split}
 \end{equation}
 where all derivatives are taken with respect to the stimulus $\theta$.
 
 Intuitively, $\IFF$  and $\IFcov$ represent the contribution of changes in
the firing rate and covariance,
respectively, to the Fisher information.  While $\IFF$ has been studied
previously, $\IFcov$ has only been examined for stimulus independent
correlation coefficients, \emph{i.e.} when $\rho' =
0$~\cite{abbott99,sompolinsky01,shamir04,shamir01}.
We separate the influence of stimulus dependent changes in correlation on
 $I_F$ as follows:

\begin{itemize}

\item {\it Correlation Coding}.  The last of the five terms in the
sum~\eqref{E:exact} is only present when $\rho' \neq 0$, and captures
the amount of information directly due to changes in
correlations~\cite{vaadia95,chacron08}.  We refer to terms in $I_F$ that are nonzero only when $\rho' \neq 0$ as the contribution of  {\it correlation coding.}  If
$f_1'=f_2'=v_1'=v_2'=0$ then all information is due to correlation coding.  It is necessary to use a nonlinear readout (decoding) scheme to recover this information~\cite{shamir04}. 

\item {\it Correlation Shaping}.  $\IFF$ is 
affected significantly by the level of correlation, $\rho$.   As $(\IFF)^{-1}$
measures the error in the optimal linear estimate of the stimulus,
the impact of changes in correlation structure on $\IFF$
represents the amount by which correlations \emph{shape} the
information available
from linear readouts of the response.  We refer to this
effect as \emph{correlation
shaping}.  
\end{itemize}

This terminology anticipates the discussion of larger populations,
where we will be interested in how the \emph{spatial structure}
together with \emph{stimulus dependent changes} of $\rho_{ij}$  affect
$I_F$.     We note that stimulus-dependent correlations can also impact the information available from the variance of the neural response (see the third term in Eq.~\eqref{E:exact}).  This is another form of correlation shaping with a marginal impact in the cases we discuss.

\begin{figure} 
	\bec
\includegraphics{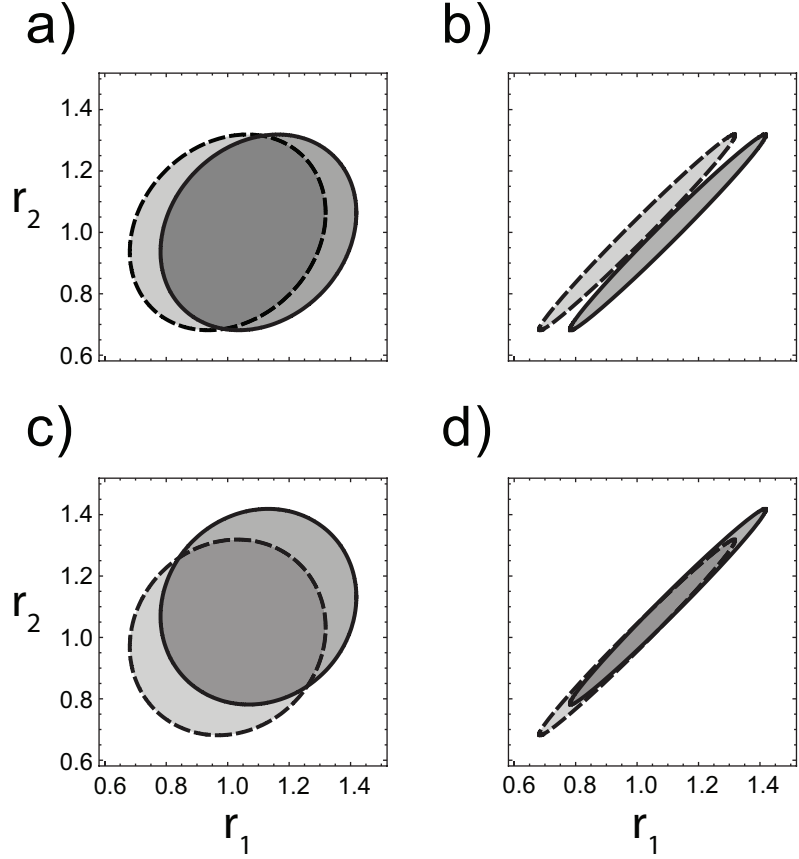}
 	\eec
\caption{ Illustration of correlation shaping for neuron pairs.  Each panel shows 50\% level curves of the joint density $p(r_1, r_2)$ in response to two nearby stimuli $\theta_A$ (dashed line) and $\theta_B$ (solid line). In all cases, $v_1 = v_2 = 1$.  A change from stimulus $\theta_A$ to $\theta_B$ 
is assumed to affect only the $f_i$, so that $\IFcov = 0$.   The beneficial effect of correlations on  $\IFF$ (first term in Eq.~\eqref{E:exact}) is illustrated in panels  a) and b).  Here $f_1'(\theta_A) \neq f_2'(\theta_A)$, and increased correlations improve discriminability.  In contrast, $f_1'(\theta_A) = f_2'(\theta_A)$ in panels
c) and d), and increased correlations reduce discriminability.  
In panels a) and b): $f_1(\theta_A) = f_2(\theta_A)= 1$, while  $f_1(\theta_B) = 1, f_2(\theta_B) = 1.1$.  In panel a), $\rho = 0.2$, while in panel b), $\rho = 0.99$.   In panels c) and d)  $f_1(\theta_A)= f_2(\theta_A) = 1$, and $f_1(\theta_B) =  f_2(\theta_B) = 1.1$.  In panel c), $\rho=0.1$, and in panel d),  $\rho=0.99$.  
} \label{f.ellipses2}
\end{figure}

We first examine the effect of correlation shaping.  A number of previous studies concluded that an increase in correlation, $\rho$, can positively impact $\IFF$ for pairs of neurons that have different ``normalized"  mean responses to the stimulus ($f_1'/\sqrt{v_1} \neq f'_2/\sqrt{v_2}$). The effect tends to be negative if the responses are similar ($f_1'/\sqrt{v_1} \approx f'_2/\sqrt{v_2}$).  Intuitively,  correlations can be used
to remove uncertainty from noisy responses of neuron pairs with differing response characteristics~\cite{oram98,averbeck06,abbott99,sompolinsky01}.

Indeed, the first term in Eq.~\eqref{E:exact}, $\left[ f_1'/\sqrt{v_1} - f'_2/\sqrt{v_2} \right]^2/(1 - \rho^2), $ increases with $\rho$, unless $f_1'/\sqrt{v_1} = f'_2/\sqrt{v_2}$.
The resulting increase in discriminability is illustrated in Fig.~\ref{f.ellipses2} where we show the bivartiate distribution $p(r_1,r_2)$ of the response to two nearby stimuli $\theta_A$ and $\theta_B$.  In panels a) and b), $v_1'=v_2'=\rho'=f_2'=0$, but $f_1' \neq 0$, so that  only the first term in $\IFF$ contributes to $I_F$.   In this example, an increase in correlation leads to a large increase in $I_F$.  In Fig.~\ref{f.ellipses2} this increase results in improved discriminability between the stimuli, \emph{i.e.} a reduction of the probability that the two stimuli will lead to the same response. However, when the two neurons respond similarly to the stimulus,  $f_1'/\sqrt{v_1} \approx f'_2/\sqrt{v_2}$,  the second term, $ 2\left[f_1' f_2'/\sqrt{v_1 v_2}\right]/(1+\rho)$, dominates.  An increase in correlations leads to a decrease in $\IFF$~\cite{sompolinsky01,averbeck06} which is reflected in  decreased discriminability between the stimuli (See panels c) and d) of Fig.~\ref{f.ellipses2}).    High values of the correlation coefficients have been used in Fig.~\ref{f.ellipses2} for easier visualization.
    
In contrast, correlation coding typically has a small effect  in the
case of two neurons, as the term $\IFcov$ is far smaller than $\IFF$.  There are two reasons for this:  The first holds only in the small correlation regime. Note that $\rho$ enters $\IFcov$ at ${\cal{O}}(\rho^2)$, while it enters $\IFF$ at ${\cal{O}}(\rho)$.  The second holds for a larger range of correlation strengths:  $v_i'/v_i$ and $\rho'$ are typically far smaller than $f_i'/\sqrt{v_i}$ and, as a result\footnote{In detail:  if responses are given by counting spikes over $\sim 1$ second, then typically $f$ takes values substantially greater than 1.  If firing is Poisson-like, then $v \approx f$.  This leads to the stated dominance of $f'/\sqrt{v}$ among these terms.}, $\IFcov \ll \IFF$. 
Therefore, under fairly general assumptions, the dominant effect of correlations on Fisher information for cell pairs is via \emph{correlation shaping} of $\IFF$.

Only close to perfect correlation, where $
\rho \approx 1$, is the impact of correlation coding potentially significant. Assuming
that $\rho' = O(1)$ as $\rho$ approaches 1, and letting $\epsilon = 1 - \rho^2$, we have
$
I_F = 2\epsilon^{-2} (\rho')^2  + O(\epsilon^{-1}).
$
Therefore, when $\rho$ is close to 1,  most  information about
a stimulus can be carried by correlation changes.  The
balance between $\IFF$ and  $\IFr$ close to perfect correlations strongly
depends on the behavior of $\rho'$ as $\rho$ approaches 1. If
$\rho'$ approaches 0 as $\rho$ approaches 1, as in~\cite{rocha06}, $\IFF$ may continue to dominate.

\paragraph{Fisher information in small  populations} Many of these observations  extend to small populations of neurons with low correlations. Let
$$
(I_F)_{i}^{\text{ mean}}  = \frac{(f_i')^2}{v_i},   \qquad
 (I_F)_{i}^{\text {var}}  =  \frac12  \left( \frac{v'_i}{v_i}\right)^2,  \qquad
 \text{and} \qquad
(I_F)_{i,j}^{\text {corr}}  = \rtwo \rtwo' \left(
\left[\frac{\rtwo'}{\rtwo} - \left( \frac{v'_i}{v_i} +  \frac{v'_j}{v_j}\right)\right] \right).
$$
We show in Appendix~\ref{A:smallrhopop} that
\begin{equation}  \label{E:smallrhopop}
\begin{split}
I_F &=  \underbrace{\sum_i   (I_F)_{i}^{\text{mean}}  - \sum_{\substack{i,j \\ i \neq j}}
\frac{f_i' f_j' \rtwo}{\sqrt{v_i v_j}} + \sum_{\substack{i,j,k \\ k \neq i,j}} \frac{f_i' f_j' \rho_{i,k}\rho_{k,j}}{\sqrt{v_i v_j}} }_{\IFF} + \\
 &\underbrace{\sum_i  (I_F)_{i}^{\text {var}} + \sum_{\substack{i,j \\ i \neq j}} \frac{\rtwo^2}{8}
 \left(\frac{v'_i}{v_i} - \frac{v'_j}{v_j}  \right)^2 + 
\sum_{i < j} (I_F)_{i,j}^{\text {corr}}}_{\IFcov} + O(\rtwo^3)  \\
\end{split}
\end{equation}
Here  $(I_F)_{i}^{\text{mean}}$ and $(I_F)_{i}^{\text {var}}$ are $O(1)$, while 
$(I_F)_{i,j}^{\text {corr}}$ is $O(\rtwo^2)$.  Therefore, $I_F$  is   a sum of contributions from individual neuron responses ($(I_F)_{i}^{\text{mean}}$ and $(I_F)_{i}^{\text {var}}$) and corrections of higher order in $\rho$ due to correlations in the response.  

Only the term  $-\sum_{i,j, i \neq j}
f_i' f_j' \rtwo / \sqrt{v_i v_j}$ in $\IFF$ is of first order in $\rho$.  This term therefore dominates the correction when correlations are small to intermediate.  In this case, correlations between differently tuned neurons again increase $I_F$, and those between similarly tuned neurons decrease $I_F.$    If correlations $\rho_{i,j}$  across the (small) population are stronger between neurons $i$ and $j$ for which $f_i'$ and $f_j'$ have opposite signs and weaker when these signs are the same, they increase $I_F$.   This is in agreement with the  
two cell case discussed above, as well as previous results~\cite{AverbeckL03,romo03,averbeck06,sompolinsky01}.

 Eq.~\eqref{E:smallrhopop} is  general, under the assumption that the response follows a multivariate Gaussian distribution.  However, the 
approximation  starts breaking down  when  $N$ exceeds $1/\rtwo$ (See Appendix~\ref{A:smallrhopop}, and Fig.~\ref{F:check}.)

\section{Large populations with no spatial  correlation decay}
\label{S:arbitrary}

In general, for large populations it  is difficult to obtain a closed form expression for
$I_F$ in terms of the variances,  correlation coefficients and firing rates.
Results are available under different simplifying assumptions
that make the problem mathematically tractable~\cite{abbott99,wilke02,shamir04}.  
In most  cases it was assumed that correlation coefficients,
$\rij$, are independent of the stimulus $\theta$, so that $\rij' = 0$.  
In the following we refer to this as the Stimulus Independent (SI) case, and 
contrast it to the Stimulus Dependent (SD) case.
The assumption that we make is that correlations between cell pairs, $\rij$, are given by
Eq.~\eqref{E:rij}, and that stimulus dependence of correlations, $S_{i,j}(\theta)$ 
takes the product form in Eq.~\eqref{e.rhoform}.

In this section we let  $c(\phi_i - \phi_j) = 1$.  Therefore,  the correlation structure
is completely determined by the stimulus.  In this case 
an analytical expression for ${\bf Q}^{-1}$ and $I_F$ can be found using
the Sherman-Morrison Formula~\cite[p.~124]{meyer2000}.  We derive the
exact expression for $I_F$ for arbitrary population sizes $N$, arbitrary response characteristics $v_i(\theta)$, $f_i(\theta)$, and $s_i(\theta)$, as well as an approximation valid for large 
populations,  in
Appendices~\ref{A:product} and~\ref{A:asymptotic}.  

To give concrete examples of how stimulus dependence of correlations impacts $I_F$ in large populations, in the remainder of the paper we further assume (as in, e.g.,~\cite{seung93,shamir01,sompolinsky01,butts06}), that cell responses follow 
tuning curves that differ only by a phase shift, so that we can write 
\begin{equation} \label{E:curves} 
f_i(\theta) = f(\theta - \phi_i), \qquad  v_i(\theta) = v(\theta - \phi_i), \qquad \text{and} \qquad s_i(\theta) = s(\theta - \phi_i),
\end{equation}
where $\theta,\phi_i \in   [0, 2\pi)$. We take all functions to be periodic.
The response, $f_i(\theta)$, is chosen so that neuron $i$ responds preferentially (with maximum rate)
to stimulus $\theta=\phi_i $, where $\phi_i$ is fixed. 
These are common assumptions that simplify the analysis considerably~\cite{sompolinsky01,wilke02}.  
Correlations are therefore determined by $\rho_{ij}(\theta)=s(\theta-\phi_i) s(\theta-\phi_j)$.

Assuming the neurons sample the stimulus space uniformly
and sufficiently densely,  we can use the continuum limit to approximate $I_F$.  In this case, an arbitrary vector ${\bf a}(\tht)$ with components $a(\tht-\phi_i)$ tends to a function $a(\theta)$ of the stimulus $\theta $.   
As we show in Appendix~\ref{A:asymptotic}, $I_F$ can then be approximated as the sum of 
\begin{equation} \label{E:asymp}
\IFF \sim  D\left(\frac{f'(\tht)}{\sqrt{v(\tht)}},  s(\tht)\right), \qquad 
\text{and}
\qquad 
\IFcov \sim \frac{N}{ \pi} \int_0^{2\pi} \left( \frac{v'(\phi)}{2v(\phi)} -\frac{s'(\phi)s(\phi)}{1 - s^2(\phi)}\right)^2 d\phi + D(G(\tht) s(\tht),  s(\tht)),
\end{equation}
where $G(\theta) = \left(s'(\theta) + \frac{v'(\theta)}{2v(\theta)} s(\theta)\right)$ and
\begin{equation} \label{E:continuumlimit}
D( a(\tht), s(\tht)) \approx \frac{N}{2\pi} \left[\int_0^{2 \pi} \frac{a^2(\phi)}{1-s^2(\phi)} d\phi - \int_0^{2 \pi} \frac{a(\phi) s(\phi)}{1-s^2(\phi)} d\phi \left/ \int_0^{2 \pi} \frac{s^2(\phi)}{1-s^2(\phi)} d\phi \right]. \right.
\end{equation}

By symmetry, neither $\IFF$, $\IFcov$ nor $I_F$ depend on $\tht$ in the large 
population limit, since the response provides equal information about any stimulus.
Therefore,  we fix $\theta=\pi$ in the following, and write the firing rates, variances and correlations 
as functions of the neurons' preferred stimuli, $\phi$.  The correlation  between two neurons
with preferred stimuli $\phi$ and $\phi'$  will  be denoted 
by $\rho(\phi,\phi')$, and $\rho(\phi) = \rho(\phi, \phi) = s^2(\phi)$ 
will be the correlation coefficient between two neurons with equal stimulus preference.

In the remainder of the paper, we make one final assumption:  that the functions $f$, $v$, and $s$ are even (i.e., symmetric around preferred orientations), as in, e.g.,~\cite{sompolinsky01,wilke02} and many other studies.  

\begin{figure} 
\begin{center}
\includegraphics[width=14cm]{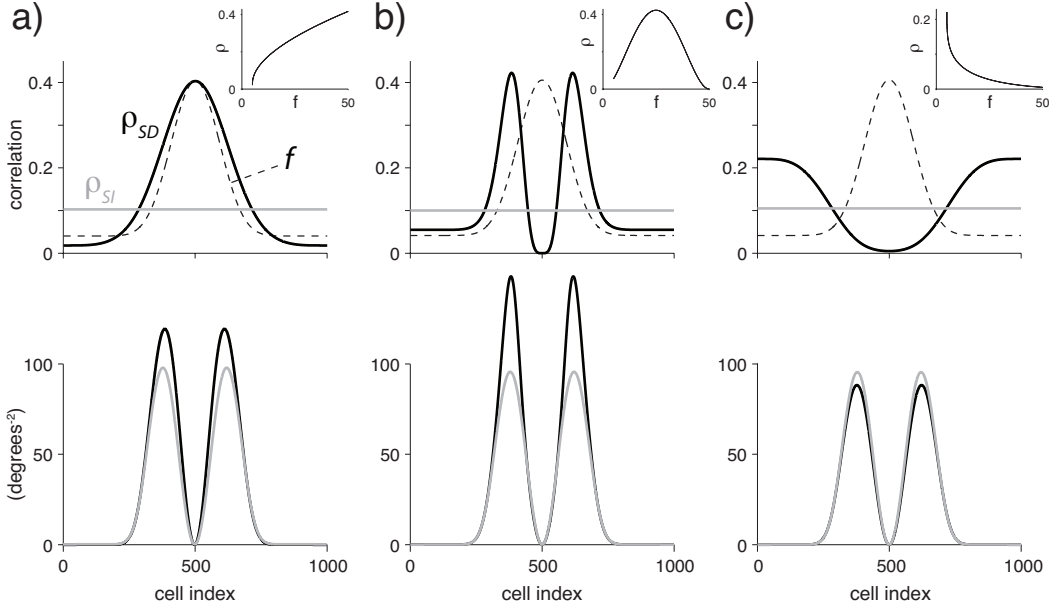}  
\end{center}
\caption[figure 2]{Examples of different correlation tuning curves and their impact on $\IFF$ for large populations. Top panels show the correlation tuning curves, $\rho(\phi) = s^2(\phi)$ for the SD (black) and SI (gray) cases  along with the (normalized) mean response $f(\phi)$ (dashed).  Average correlations are matched to equal 0.1 in all cases. Insets illustrate the $\rho-f$ relationship for each choice of the correlation tuning. Bottom panels show the integrand of Eq. (\ref{E:IFFcontinuum}) for the SD (black) and SI (gray) cases. {\bf a)}  $\rho-f$ follows a concave increasing curve, and $\rho(\phi)$ shows a slightly broader tuning than $f(\phi)$ in the SD case, resulting in  a substantial increase in $\IFF$  with respect to the SI case (increase of  $\sim 21\%$).  {\bf b)}  $\rho-f$ is non-monotonic, and $\rho(\phi)$ is bimodal and matches $(f'(\phi))^2/v(\phi)$ in the SD case. This yields a larger enhancement of $\IFF$ with respect to the SI case (increase $\sim29\%$). {\bf c)} Correlations that decrease with rate have a negative impact on $\IFF$ (decrease of $\sim 7\%$ compared to the SI case). In all cases $\IFF$ was computed in the large $N$ limit using Eq.~\eqref{E:asymp}. Parameters: average correlation coefficient ${\bar s}^2= 0.1$ in all cases (larger values, e.g. 0.2, will typically more than double the difference in $\IFF$ between SD and SI cases). In all cases $f(\phi)=5 + 45 a^6(\phi)$ with $a(\phi) = 1/2 (1 - \cos(\phi))$, and $v(\phi)=f(\phi)$ (Poisson).  ({\bf a}) $s(\phi)$ =$ k_\rho + b_\rho a^2(\phi)$ where $k_\rho=0.135$ and $b_\rho=0.5$; ({\bf b}) $s(\phi) = 4r_{\text{max}} f(\tht) [f_{\text{max}}-f(\tht)]/f_{\text{max}}^2 $ with $r_{\text{max}} = 0.65$ and $ f_{\text{max}} = 50$;  ({\bf c}) $s(\phi)$ =$ k_\rho + b_\rho a^2(\phi)$ where $k_\rho=0.47$ and $b_\rho=-0.4$.  (See Appendix E.)}
\label{F:example}
\end{figure}

\paragraph{\bf Effects of stimulus-dependent correlations on $\IFF$:} 

To illustrate how stimulus dependence of correlations can influence the 
information contained in the population response we first consider $\IFF$.
Even  when  correlations are small, this stimulus dependence can have a strong effect via correlation shaping.  

Since $f(\phi)$ and $v(\phi)$ are  even,  $f'(\phi)/\sqrt{v(\phi)}$ is
odd.  Therefore, setting $a(\phi) =  f'(\phi)/\sqrt{v(\phi)}$, the second term in Eq.~\eqref{E:continuumlimit} vanishes, and
\begin{equation} \label{E:IFFcontinuum}
\IFF =  \frac{N}{2\pi} \int_0^{2 \pi} \frac{(f'(\phi))^2}{v(\phi)}\frac{1}{1-s^2(\phi)} d\phi.
\end{equation}
Although $\IFF$ is the average of the Fisher information $[f_i']^2/v_i$ of {\it single} neurons, with a weighting factor, caution needs to be exercised when interpreting this result.  Eq.~\eqref{E:IFFcontinuum} is the result of simplifying an expression derived from all pairwise interactions across the population.

\begin{figure}
\hspace{-.87in}    
\bec
\includegraphics[width=.75 \linewidth]{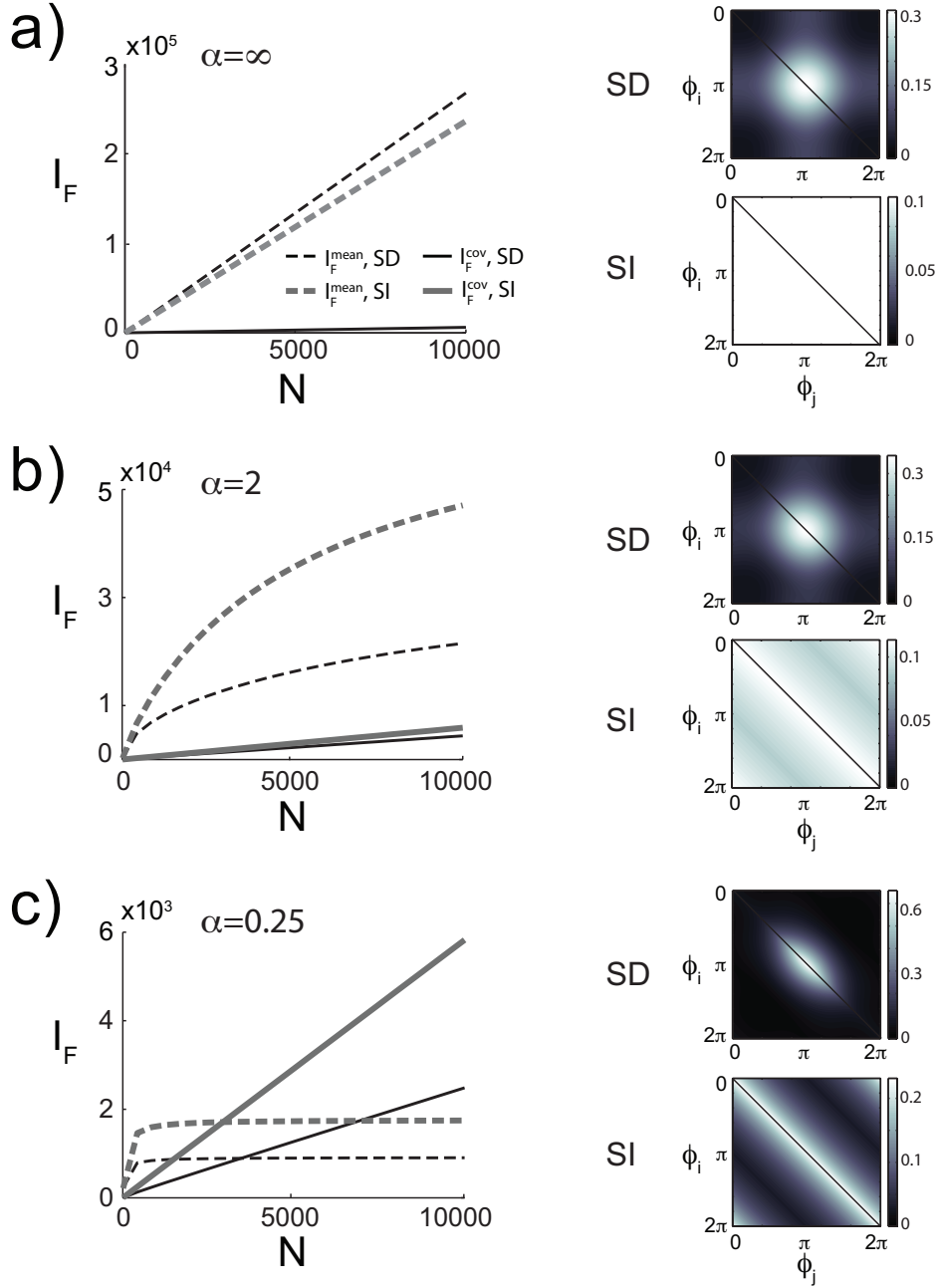}
\eec   
\caption{$\IFF$ and $\IFcov$ as a function of population size $N$, for matched SD and SI correlation cases and various correlation decay lengthscales.  Here, correlation is assumed to increase with firing rate,  as in  Fig.~\ref{F:example}a).  The coefficients $k_{\rho}$ and $b_{\rho}$ defining $s(\phi) = k_{\rho}+b_{\rho}a^2(\phi)$ are chosen   to keep the average correlation coefficients over the population equal to 0.1 (See Appendix~\ref{A:constantcorrelations}).   The corresponding correlation matrices, $\rho_{i,j}$, for the SD and SI cases are also shown (on-diagonal terms are set to 0 in these plots).}
\label{F:nullfull1}
\end{figure}

In the SI case, each $s_i(\theta)$ %
is independent of the stimulus, and $s(\phi)$ is therefore constant across the population:  $s(\phi) = \bar s$. We focus on comparisons between SI and SD cases matched to have the {\it same} average correlation coefficient across the population.    {\it We therefore assess the effects of the  stimulus-dependence of correlation, as opposed to the level of correlations.}  Specifically, we ensure that the average correlation coefficient across the population in the SD case,   $(4\pi^2)^{-1}\int_0^{2\pi} \int_0^{2\pi}   s(\phi_1)s(\phi_2) \, d\phi_1 \, d\phi_2$,  equals that in the SI case by setting $\bar s=1/(2\pi)\int_0^{2\pi}s(\phi)d\phi$.  Examples of typical matched correlation matrices, $\rho_{ij}$, in the SD and SI cases, are shown in the right hand column of Fig.~\ref{F:nullfull1}.

Panels a) and b) of Fig.~\ref{F:example} illustrate how correlation shaping may increase $\IFF$ in the SD case over  the SI case.  In each, stimulus-dependence of correlations arises from a different relationship between stimulus-induced firing rate and correlation (see insets).   In a), $\rho(\phi)$ increases with $f(\phi)$, as in~\cite{rocha06} and certain regimes in~\cite{Bin+01,kohn05,greenberg08}.  In b), $\rho(\phi)$ first increases with $f(\phi)$, and then decreases, as in feed-forward networks with refractory effects~\cite{brown07}.  Importantly, for both panels a) and b), correlations are high between neurons that individually carry most information about the stimulus (i.e., between neurons with large values of $(f'(\phi))^2/v(\phi)$).  Therefore, the weighting factor $1/(1-s^2(\phi))$  assigns a greater contribution of these more-informative cells to the weighted average in Eq.~\eqref{E:IFFcontinuum} for the SD case, leading to the increase in $\IFF$.  

On the other hand, panel c) of Fig.~\ref{F:example} illustrates a case in which correlations decrease
with firing rates, as observed in~\cite{AksayBST03}.  As a result, correlations between the most informative neurons are smaller than average, and correlation shaping negatively impacts $I_F$.  We note that in all panels maximum pairwise correlations  satisfy $\rho_{\text{max}} \lesssim 0.45$, within the range typically reported~(e.g., \cite{gutnisky08,roelfsema08,zohary94}).   Increasing this maximum, without changing the mean correlation, can make these correlation shaping effects more pronounced.

A different way of seeing how  $\IFF$ can be greater in the SD than the SI case is given in Fig.~\ref{F:nullfull1}a.  Here, $\IFF$ and $\IFcov$ are computed numerically, and plotted as a function of the population size $N$, for both the SD and SI cases that correspond to the example of correlations increasing with rate (Fig.~\ref{F:example}a).  Note that $\IFF$ dominates $\IFcov$ over a wide range of $N$ and that the total Fisher information, not just $\IFF$, is greater in the SD vs. SI case.  Moreover, the continuum limit given in Eq.~\eqref{E:IFFcontinuum} appears valid even for moderate population sizes.

\begin{figure}[h!]
	\begin{center}
\includegraphics[width=.75\linewidth]{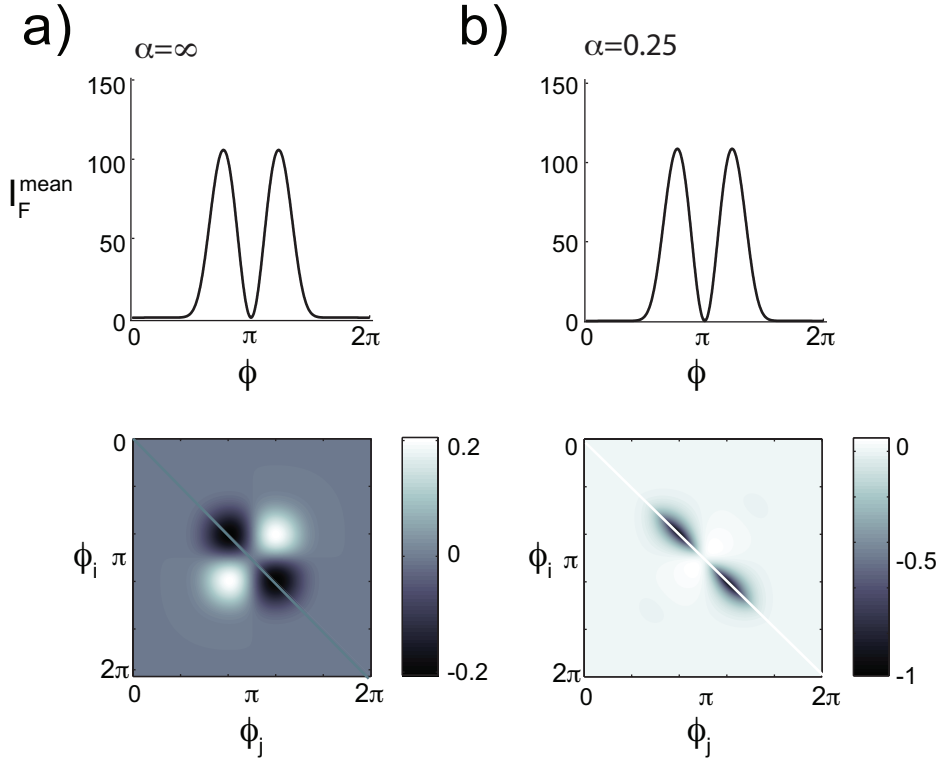}	
\end{center}
\caption{Plots 
of the matrix $f_i' f_j' Q^{-1}_{i,j}$ whose double sum determines $\IFF$. {\bf a)} no spatial correlation decay,  and {\bf b)} spatial decay with $\alpha=0.25$.  Top:  on-diagonal terms of matrix.  Bottom:  off-diagonal terms (with on-diagonal values set to 0 for ease of visualization). }
\label{F:FImatrix}
\end{figure}

Care needs to be taken when trying to intuitively understand these population-level effects of stimulus-dependent correlations on $\IFF$ by invoking the case
of two neurons studied in Section~\ref{s.twocells}.  Consider the case of correlations increasing with firing rate~(Figs.~\ref{F:nullfull1}a, \ref{F:FImatrix}a). As  noted in the discussion of 
Eq.~\eqref{E:exact}, an increase in correlations between two similarly tuned neurons will typically have a negative
impact on $\IFF$, due to the dominance of the second term of $\IFF$ in Eq.~\eqref{E:exact}.  On the other hand, Eq.~\eqref{E:IFFcontinuum} shows that increasing correlations
between the most informative neurons in a large population, regardless of the similarity of their tuning, has a positive impact.  The two results are not contradictory.  Consider the pairwise sum of the two-neuron $\IFF$ from Eq.~\eqref{E:exact} over \emph{all neuron pairs} in the population.  Note that the second term of $\IFF$ in Eq.~\eqref{E:exact} can be expected to be matched with one of equal and opposite sign in  such a sum, if the 
tuning curves are symmetric, and correlations depend only on firing rate.  Therefore, the typically-dominant second term cancels, and it is the 
{\it first} term in Eq.~\eqref{E:exact}, always positively impacted by the presence of correlation, that remains.   Moreover, examination of this first term in Eq.~\eqref{E:exact} {\it} does show similarity with Eq.~\eqref{E:IFFcontinuum}: in both cases, assigning largest correlations $\rho_{i,j}$ or $s(\phi,\phi')$ to most-informative neurons will yield the greatest total value of $\IFF$.

Fig.~\ref{F:FImatrix}a) shows that this cancellation argument, while not {\it directly} applicable, is at least analogous to what happens when computing $\IFF$ for the large population via the complete expression~\eqref{E:asymp}.  The sum of the terms $f_i' f_j' Q^{-1}_{i,j}$ 
defines the linear Fisher information, $\IFF = \sum_{i,j} f_i' f_j' Q^{-1}_{i,j}$ (see Eq.~\eqref{E:IFFgeneral}).  Under the present symmetry assumptions, the off-diagonal
terms cancel, and only the diagonal terms contribute to the sum.
In Appendix~\ref{A:asymptotic} we show that $Q^{-1}_{i,i} = [v_i (1- s_i^2)]^{-1}$, 
in agreement with the remaining term in Eq.~\eqref{E:IFFcontinuum}. 

These observations are robust to the presence of weak asymmetry in the functions $f$, $v$, and $s$.   For instance, when 
the tuning curve $f(\tht)$ is  a sum of a symmetric and small asymmetric part, $f_{\text{sym}}(\tht) + \epsilon f_{\text{asym}}(\tht)$, an examination of  Eq.~\eqref{E:continuumlimit} shows
that the impact of the asymmetry on $\IFF = D(\frac{f'(\tht)}{\sqrt{v(\tht)}},s(\tht))$ is of 
order $O(\epsilon N)$, while  $\IFF$ is $O(N)$.   However, we show in the next section that the large population
limit can be changed significantly when $c(\phi_i - \phi_j)$ is not constant.

\paragraph{\bf Effects of stimulus-dependent correlations on $\IFcov$:} 
Having discussed $\IFF$, we now turn to the impact on $\IFcov$ of the stimulus-dependence of correlations.  In Appendix~\ref{s.ifcov} we show that this impact is negligible for small to intermediate correlations, and that
	\beq \label{E:IFcovcontinuum}
\IFcov \approx \frac{N}{2 \pi} \int_0^{2\pi} \frac{v(\phi)}{2v'(\phi)} d\phi.
	\eeq
Moreover, as discussed in Section~\ref{s.twocells}, values of ${v(\phi)}{2v'(\phi)}$ are typically smaller in magnitude than values of $\frac{(f'(\phi))^2}{v(\phi)}$.  Therefore, for small to intermediate correlations the major contribution of the stimulus-dependence of correlations comes from
$\IFF$ rather than $\IFcov$.  This agrees with the case of two cells (Sec.~\ref{s.twocells}).   Asymptotic estimates of the integrals in $\IFcov$ show that this remains true even for correlation coefficients {\it close to one}.   The dominance of $\IFF$ over $\IFcov$  is apparent in Fig.~\ref{F:nullfull1}a).  As we show in the next section, however, that this dominance may no longer hold in the 
presence of spatial decay of correlations~\cite{sompolinsky01,shamir04}.  

\medskip

\noindent{\bf Summary of Sec.~\ref{S:arbitrary}:  }Stimulus-dependence may shape the structure of correlations  so that  neurons that are most informative about the stimulus  presented are most highly correlated.  This can lead to an increase in overall information.    This is possible even when the average correlations across the population are low, but not when correlations are fixed, or if all neurons have identical mean responses.

\section{Effects of correlation stimulus dependence  in the presence of spatial decay}
\label{s.decay}

In this section we examine how stimulus-dependent correlations  affect $I_F$ in the presence of spatial correlation decay.  We again assume that correlations and rates are described by Eqs.~(\ref{E:rij}--\ref{e.rhoform}), but we now assume that
$$
c (\phi_i - \phi_j) = C  \exp \left[ -  \frac{|\phi_i - \phi_j|}{\alpha} \right].
$$
The constant $\alpha$ determines the spatial range of correlations, while $C$ was chosen so that the 
average correlation across the population $\langle \rho_{i,j} \rangle$ remains constant as other parameters are varied
(for details see Appendix~\ref{A:constantcorrelations}). As an exact
expression for the inverse of the covariance matrix is difficult to obtain, we 
study this case numerically, and give an intuitive explanation of the results. 

\paragraph{\bf Effect of correlation shaping on $\IFF$:}  When $\alpha = \infty$, there is no spatial decay,  and we 
are in the situation discussed in the previous section: $I_F$ is typically dominated by $\IFF$, which grows linearly with population size $N$ (Fig.~\ref{F:nullfull1}a).  However, for finite values of $\alpha$, $\IFF$ generally {\it saturates} with increasing $N$ (Fig.~\ref{F:nullfull1}b,c).  This agrees with earlier findings for stimulus-independent correlations~\cite{shamir04,shamir06}.  

Additionally, effects of stimulus-dependence in correlations on $\IFF$ can be {\it reversed} for finite values of $\alpha.$  For example, assume  that $s_i(\phi)$ increases with the firing rate, as in  Fig.~\ref{F:example}a).  When $\alpha=\infty$, stimulus-dependence of correlations increases $\IFF$ (Fig.~\ref{F:nullfull1}a).  However,  for finite $\alpha$, this stimulus dependence has a negative impact on $\IFF$ (Fig.~\ref{F:nullfull1}b,c).  

Intuitively, this may be due to spatial correlation decay reducing 
correlations between  neurons with \emph{differing} stimulus preferences.  The 
negative impact of correlations between similarly tuned neurons on $I_F$ is no
longer balanced by the positive impact on differently tuned neurons.  Indeed, the stronger
the spatial decay of correlations, the more this balance is broken.  Therefore, the
cancellation arguments presented in the previous section no longer hold -- compare Fig.~\ref{F:nullfull1}b) and c) -- and it is no longer the case that simply increasing correlations for more-informative neurons will increase $\IFF$.    Instead, correlation structures that increase correlation for similarly vs. differently tuned neurons can again be expected to decrease $\IFF$.  Figure~\ref{F:FImatrix} shows that this is the precisely the effect of the SD vs. SI correlation structures.

\begin{figure} 
\hspace{-.45 in}
	\bec
\includegraphics[width=.9 \linewidth]{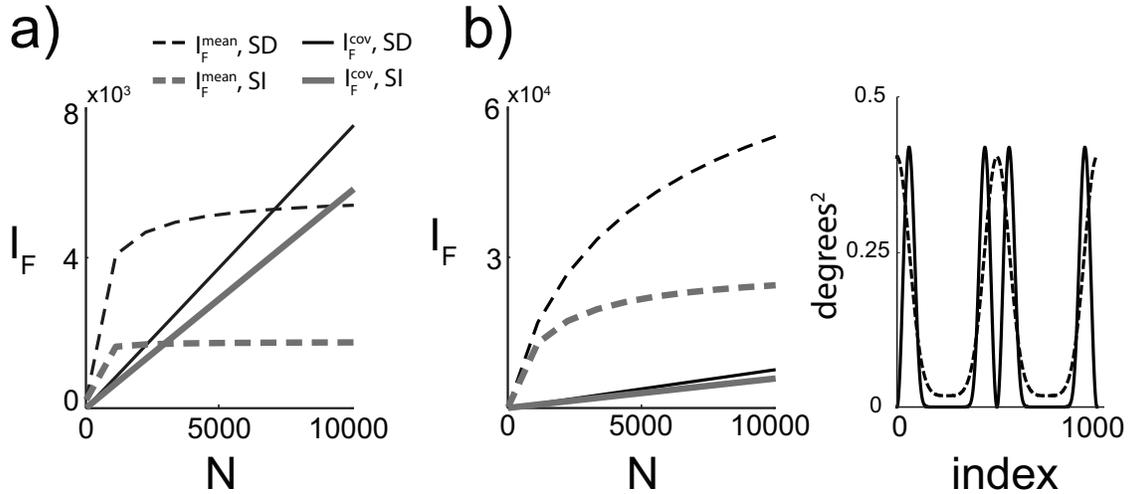}
	\eec
\caption{Examples where the Fisher information is larger in the SD than the SI case, despite strong correlation decay $\alpha=0.25$.  In each case (as in all of Figs. 2-6), $s(\phi)$ is set so that the average correlation coefficient $\rho$ across the population is 0.1.   {\bf a)} Correlations {{\bf \it decrease}} with firing rate, as in Fig.~\ref{F:example}c). {\bf b)} The response of a population with two subpopulations each tuned to the stimulus.   In the right panel, as in Fig.~\ref{F:example}, $(f'(\phi))^2/v(\phi)$ is scaled and represented by the solid line, while 
 $\rho(\phi) = s(\phi)^2$ is represented by the dotted line. The effects of spatial correlation decay are not shown. The response of each cell in the population follows a unimodal tuning curve; however,
there are two sets of cells at different spatial locations, that share the same stimulus preference.  The left panel shows the effect of this arrangement on the Fisher information.
Other parameters are as in Fig.~\ref{F:nullfull1}.    }
\label{F:nullfull2}
\end{figure}

As a second example, assume that correlations decrease, rather than increase, with firing rate, as in Fig.~\ref{F:example}b. In this case, 
correlations between similarly tuned, strongly responding neurons are decreased.  As expected from the arguments above, stimulus-dependent correlations then increase $\IFF$ over its value in the stimulus-independent case.  Moreover, absolute levels of $I_F$ increase twofold compared to the analogous case where correlations increase with rate (compare Fig.~\ref{F:nullfull1}c
and Fig.~\ref{F:nullfull2}a).  
 
However, in all of these cases, note that levels of $I_F$ are lower in the presence of correlation decay for {\it{both}} SD and SI cases.  We now mention one way in which this can be mitigated.  As illustrated in Fig.~\ref{F:nullfull2}b), we increase the number of areas or subpopulations that respond strongly to a given stimulus.   The response of each cell still follows a unimodal tuning curve, as above.  However, the entire population has a number of cells at different spatial locations that share the same stimulus preference.    Therefore, cells in different subpopulations are only weakly correlated and can be thought of as members of different, nearly independent populations.  As Fig.~\ref{F:nullfull2}b) shows, this boosts overall levels of $\IFF$, while maintaining the benefit of stimulus-dependence in correlations within individual subpopulations.  


In sum, the spatial decay of correlations has a strong negative effect on linear
Fisher information $\IFF$.  If correlations depend on stimuli via an increasing relationship with firing rate, this effect can be accentuated, with levels of $\IFF$ decreasing by a further factor of two for SD vs. SI cases.  However, the opposite effect occurs if correlations decrease with rate:  stimulus-dependence can then approximately double $\IFF$. 

\paragraph{Effect of correlation coding on $\IFcov$:}  For large populations,  Figs.~\ref{F:nullfull1}, \ref{F:nullfull2} show that information
can be carried predominantly by $\IFcov$, and this dominance 
is more pronounced as the correlation lengthscale $\alpha$ decreases.  This agrees with earlier findings~\cite{sompolinsky01,shamir04}.   Moreover, we see that the effects of stimulus dependence of correlations on $\IFcov$ have the same ``sign" as those on $\IFF.$  Specifically, when correlations increase with rate, as in Fig.~\ref{F:nullfull1}, both $\IFF$ and $\IFcov$ are lower in the SD than in the SI cases, for finite values of correlation length $\al$.  Also, when correlations decrease with rate, as in Fig.~\ref{F:nullfull2}, corresponding values of both $\IFF$ and $\IFcov$ are higher for the SD than the SI case.   

The effects of stimulus dependence on the (dominant) $\IFcov$ terms can be attributed to {\it correlation coding}.  In detail, the contribution of $\rho'
_{ij}(\tht)$ terms to $\IFcov$ can be isolated numerically by simply computing $\IFcov$ twice:  once with these terms at the nonzero values expected from stimulus dependence, and once after ``artificially" setting all of these terms equal to zero.  The difference is the contribution to $I_F$ attributable directly to changes in correlation with the stimulus (i.e., correlation coding, as opposed to the correlation shaping effects that have been the focus of much of the previous discussion).  Our calculations (not shown) indicate that almost the entire increase, or decrease, of $\IFcov$ in the SD relative to the SI cases is due to this correlation coding.

\paragraph{Remark 1:}
More heterogeneous populations of neurons have been shown to yield higher values of Fisher information in some cases 
~\cite{shamir06,chelaru08}. 
We modeled such heterogeneity by randomly and independently jittering the 
tuning curves of the neurons, while preserving the expected correlation
between pairs.   Perturbing the different tuning curves by 10\% had 
a relatively small impact on the present results.  Specifically, $I_F$ terms still increased (or decreased) in the same SD vs SI cases.  Moreover, although $\IFF$ does not necessarily
saturate, for small perturbations $\IFcov$ still dominates even at large population sizes.  

\paragraph{Remark 2:} As discussed in Section~\ref{S:fisher}, it has been observed that correlations between neuronal responses decrease
with the difference between their preferred stimuli~\cite{zohary94,lee98}.  
This effect can also follow from stimulus-dependence of correlations:  When correlations increase with firing rate,  two neurons that both respond strongly to similar stimuli will be more correlated than 
those of neurons whose preferences differ.   As neurons with similar preferences in stimuli
can be expected to be physically closer in the cortex, stimulus dependence 
can result in correlations  that decay  
with physical distance~\cite{brown07}.  This is quite different from the case where  
physically distant cells are less correlated due to a smaller overlap in their inputs.
With stimulus-dependence of correlations, two distant cells, one or both of which
are responding strongly, may be more correlated than two nearby
cells that are both responding weakly (see Fig.~\ref{F:nullfull1}).

\section{Discussion}

Correlations in the neural response have the potential to both positively and negatively impact the ability of a population to carry information about stimuli.  Intuitively, correlated fluctuations imply a common component in the  response noise of different neurons.  Similarly tuned neurons may provide redundant information, as  the common noise component cannot be directly averaged away~\cite{johnson80,britten92,zohary94}.  However, it is also possible that noise can be removed by taking differences between neural responses~\cite{abbott99}.  The net effect of correlations on population level information therefore depends on the balance among different effects.  

We considered neuronal populations with stimulus-dependent correlations  
and discussed two ways in which such stimulus dependence influences Fisher information.  The first, {correlation coding}, refers to the information directly carried by changes in correlation structure in response to stimuli.  The second, {correlation shaping}, refers to the impact of stimulus dependence on  information carried by the mean and variance of neural responses.  In different cases, we derived expressions for the Fisher information that isolate correlation shaping and correlation coding effects: For cell pairs, and small-to-intermediate populations Eqs.~(\ref{E:exact}), (\ref{E:smallrhopop}) are valid for general correlation structures. For correlations with product structure, $\rho_{ij}(\tht)=s_i(\tht) s_j(\tht)$, expressions  are derived for populations of arbitrary size $N$, with simplifications in the continuum limit $N \rightarrow \infty$ (Eqs. \eqref{E:IFFcontinuum} and \eqref{E:IFcovcontinuum}).	

These expressions allow us to make a number of general observations.  
For typical firing regimes, we find that the effects of correlation shaping dominate over those of correlation coding for pairs of neurons or small populations with weak-to-moderate correlations, with most information being carried by $\IFF$.  Correlation coding only becomes significant for strong correlations.  However, for large populations the answer is different.  Without spatial decay of correlations, correlation shaping and $\IFF$ dominate (cf.~\cite{shamir04,shamir06}) regardless of correlation strength.  However, correlation coding and $\IFcov$ become important in the presence of decay.

Additionally, for pairs of neurons or small populations with weak correlations, correlated responses between similarly tuned neurons typically decrease $\IFF$, while correlations between oppositely tuned neurons increase $\IFF$, as has been shown in related settings (cf.~\cite{AverbeckL03,romo03,averbeck06,sompolinsky01}).  However, for large populations with symmetric and uniformly distributed tuning curves, the situation may be quite different. For correlations with  product structure and without spatial decay, $\rho_{ij}(\tht)=s_i(\tht) s_j(\tht)$,  correlations between the ``most-informative"  neurons
(those with largest $f_i'(\tht)/\sqrt{v_i(\tht)}$)  have the greatest 
impact on $\IFF$, regardless of similarity of tuning.   Some forms of stimulus dependence can increase these correlations,  providing a boost to the Fisher information; others decrease these correlations and hence the Fisher information.  	Interestingly, in the presence of spatial decay of correlations, these effects of stimulus dependence on Fisher information are typically reversed.  We note one interpretation:  since spatial decay tends to decrease Fisher information, the correct stimulus dependence of correlations can counterbalance this effect.

What biological mechanisms could underly different patterns of stimulus-dependent correlation?  One is the co-tuning of correlation and response rate that has been observed in feed-forward networks~\cite{rocha06,brown07}.  More complex network effects could be behind the decreasing trend of correlation with rates seen in~\cite{AksayBST03}. Moreover, stimulus-dependent adaptation of correlations has been observed in the visual cortex~\cite{kohnprep,ghisovan08,gutnisky08}.  Our study points to the potentially distinct impacts of the mechanisms on population codes.

Fisher information is only one of the possible metrics that can be used to
quantify the impact of correlations.  However, its close connection with
stimulus discriminability~\cite{dayan2001}, relative ease of computation compared
to other metrics, and recent use in experimental settings~\cite{gutnisky08,AverbeckL03}
make it a good starting point.  Future work will extend our study of the impact of correlation stimulus dependence  to other metrics, such as mutual information, adding to results of~\cite{kohn07,PanzeriSTR99}.  

Another important question for future work comes from decoding:  how can  information encoded in correlation changes be read out?  For cases in which information is dominated by $\IFF$ terms, a linear readout will suffice; however, when $\IFcov$ dominates, as for large populations with distance-dependent decay of correlations, nonlinear schemes   are required~\cite{shamir04}.

\paragraph{Acknowledgments}   We thank Bruno Averbeck, Jeff Beck, and Adam Kohn for their insights and helpful comments and suggestions.  E. S.-B. holds a Career Award at the Scientific Interface from the Burroughs-Wellcome Fund. This research was also supported NIH grant   DC005787-01A1 (J. R.), a Texas ARP/ATP, and
NSF grant DMS-0604429 to K.J., and NSF grant DMS-0817649 to B. D., K.J., and E. S.-B.

\appendix

\section{Fisher information for small populations with small correlations}
\label{A:smallrhopop}

The appendices contain a number of exact expressions and approximations
of the Fisher information for both intermediate and large populations.   These results should
be useful in the further analysis of the impact of correlations in settings similar and distinct from those studied here.

\newcommand{\trij}{\tilde{\rho}_{i,j}}
\newcommand{\e}{\epsilon}
The approximation in~\eqref{E:smallrhopop} is obtained from
the assumption $|\rij| \ll 1$.  Defining $\e \trij = \rij$, we can write
$$
Q_{i,j} = \delta_{i,j} v_i + \e ( 1- \delta_{i,j})  \trij \sqrt{v_i v_j}.
$$
Therefore ${\bf Q}$ is a perturbation of a diagonal matrix ${\bf R}$ with
entries $R_{i,j} = \delta_{i,j} v_i(x)$, and the perturbation
$\e {\bf S}$ where $S_{i,j} = ( 1- \delta_{i,j})  \trij(x) \sqrt{v_i(x) v_j(x)}$.
We can now use the standard matrix perturbation result (see
also~\cite{wilke02,demmel97})
\begin{equation} \label{E:asymptotic}
\begin{split}
{\bf Q}^{-1} & =  \left[{\bf R} ({\bf I} + \e {\bf R}^{-1}{\bf S}) \right]^{-1} =  ({\bf I} + \e {\bf R}^{-1}{\bf S})^{-1} {\bf R}^{-1}  \\
&= \left[\sum_{i = 0}^{\infty}  (- \e {\bf R}^{-1}{\bf S})^i \right] {\bf R}^{-1} \\
&= {\bf R}^{-1} - \e {\bf R}^{-1}{\bf S}{\bf R}^{-1} +
\e {\bf R}^{-1}{\bf S}{\bf R}^{-1}{\bf S}{\bf R}^{-1} + (\e^3).
\end{split}
\end{equation}
The equality on the second line holds whenever $\| \e {\bf R}^{-1}{\bf S} \| < 1$ for a norm
$\| \cdot \|$ which is consistent with itself~\cite[Lemma 2.1]{demmel97}. 
Using~\eqref{E:asymptotic}, we obtain
\begin{equation} \label{E:inverseapprox}
Q^{-1}_{i,j} = \delta_{i,j} \frac{1}{v_i} - \e ( 1- \delta_{i,j}) \frac{\trij}{\sqrt{v_i v_j}}
+ \e^2  \sum_{\substack{k \\ k\neq i,j}}  \frac{\tilde{\rho}_{i,k} \tilde{\rho}_{k,j}}{\sqrt{v_i v_j}}.
\end{equation}

Using this equation, the first term in the expression for
$I_F$, ${\bf f}^T {\bf Q}^{-1} {\bf f}$, can be computed
directly, to obtain the expression on the first line of~\eqref{E:smallrhopop}.
The second term, Tr$[({\bf Q}' {\bf Q}^{-1})^2]/2$, can
be computed similarly, through a lengthier computation.  This
computation can be simplified using the observations  in the
next section. This gives Eq.~\eqref{E:smallrhopop}, keeping terms up to second order.

The convergence of the sum on the  second line of~\eqref{E:asymptotic} is
not guaranteed if $\| \e {\bf R}^{-1}{\bf S} \| > 1$.   This implies that for fixed $\epsilon$, the approximation~\eqref{E:inverseapprox}   will
break down for sufficiently large $N$ (typically about when $N > 1/\epsilon$).

\section{General expression for $I_F$ in the product case}
\label{A:product}

In this section we use the Sherman-Morrison Formula~\cite[p. 124]{meyer2000} to derive a general expression for the Fisher information
in the product case.
Let
\begin{equation} \label{E:S}
 {\cal{S}}=\sum_{j=1}^N \frac{s_j^2}{(1 - s_j^2)}.
 \end{equation}
Then
\begin{equation} \label{E:inverseQ}
Q^{-1}_{i,j} = \begin{cases}
\displaystyle \frac{1}{v_i(1-s_i^2)} \left( 1  - \frac{s_i^2}{(1 + {\cal S})(1-s_i^2)}\right)  &  \text{ if $i = j$}  \\
& \\
\displaystyle   -\frac{s_i s_j}{\sqrt{v_i v_j}(1 +{\cal S})(1-s_i^2)(1-s_j^2)} &  \text{ if $i \neq j$}.
\end{cases}
\end{equation}

Using this equation we can obtain a compact expression for $I_F$.  The term
resulting from changes in the mean number of spikes as the stimulus varies is
given directly from definition~\eqref{E:fullfisher} as
\begin{equation} \label{E:mean1}
I_F^{mean}(x)=\sum_{i,j=1}^N f'_i f'_j Q^{-1}_{i,j}.
\end{equation}

The contribution to $I_F$ due to changes in the covariance, given by
$\IFcov=$Tr$[({\bf Q}'{\bf Q}^{-1})^2]/2$, can be expressed
compactly by introducing
\begin{equation} \label{E:SandR}
R_i = \frac{d}{dx} \ln s_i = \frac{s'_i}{s_i}, \qquad \text{and} \qquad
Z_i = \frac{d}{dx} \ln (s_i \sqrt{v_i}) = \frac{s'_i}{s_i} + \frac12 \frac{v'_i}{v_i}.
\end{equation}
Note that when $\rij$ have the form given in Eq.~\eqref{E:rij}, $c(\phi_i - \phi_j) = 1$, and the stimulus dependence of correlations, $S_{i,j}(\theta)$ 
takes the product form in Eq.~\eqref{e.rhoform}
If ,
we can write
$$
Q'_{i,j} = (Z_i + Z_j - 2 \delta_{i,j} R_i) Q_{i,j},
$$
where $Z_i$ and $R_i$ are defined in~\eqref{E:SandR}. Following this observation,
we can follow the computations in~\cite[Appendix A]{wilke02}, to obtain
\begin{equation*}
\begin{split}
\frac{\text{Tr}[({\bf Q}' {\bf Q}^{-1})^2]}{2} =  \sum_{k = 1}^{N} Z_k^2 +
 \sum_{k,l = 1}^{N} Q_{k,l} Z_k Z_l Q_{l,k}^{-1}  - 4 \sum_{k = 1}^{N}
Q_{k,k} Z_k R_k Q_{k,k}^{-1}
+  2  \sum_{k,l = 1}^{N} Q_{k,k} Q_{l,l} R_k
R_l  Q_{k,l}^{-1} Q_{l,k}^{-1}.
\end{split}
\end{equation*}
Observing that ${\bf Q}^{-1}$ is self-adjoint,
we obtain
\begin{equation} \label{E:fisher2}
\begin{split}
\IFcov &=
\sum_{i=1}^N  (Z_i)^2 \left[ 1 + Q^{-1}_{i,i}v_i  (1 - s_i^2)\right] + \sum_{i,j}^N
Z_i Z_j  s_i s_j \sqrt{v_i v_j} Q^{-1}_{i,j} \\
 &+ 2 \sum_{i,j}^N R_i R_j \left[ \sqrt{v_i v_j} Q^{-1}_{i,j} \right]^2 -
 4 \sum_{i=1}^N Z_i R_i v_i Q^{-1}_{i,i}. \
 \end{split}
\end{equation}
Therefore, $I_F$ is the sum of~\eqref{E:mean1} and~\eqref{E:fisher2}.

The contribution to $I_F$ due to only changes in the variances can be obtained from
Equation~\eqref{E:fisher2} by setting $R_i = 0$ and replacing $Z_i$ by
$v'_i/(2 v_i)$, so that
\begin{equation} \label{E:var}
I_F^{var} =
\sum_{i=1}^N  \left(\frac{v'_i}{2v_i}\right)^2 \left[ 1 + Q^{-1}_{i,i}v_i  (1 - s_i^2)\right] + \sum_{i,j}^N
 \frac{v'_i v'_j  s_i s_j}{4 \sqrt{v_i v_j}}  Q^{-1}_{i,j}.
\end{equation}
The contribution due to correlation stimulus dependence is therefore
$$
\IFr  = \IFcov - \IFV.
$$

\section{Asymptotic results}
\label{A:asymptotic}

The expression for $I_F$ derived in Appendix~\ref{A:product} can be simplified
considerably for large cell populations. 
If $N$ is large and
$0 < \epsilon < s_i < 1-\delta$ for some $\epsilon, \delta > 0$, then
 ${\cal S} = O(N)$, where ${\cal S}$ is defined in~\eqref{E:S}.  The 
 assumptions on $s_i$ are not essential, but make the derivation of the asymptotic expressions easier. 
  
Keeping only the leading order terms in~\eqref{E:inverseQ}
 we can write
\begin{equation} \label{E:inverse_asymptot}
Q^{-1}_{i,j} \approx \begin{cases}
\displaystyle \frac{1}{v_i(1-s_i^2)}   &  \text{ if $i = j$}  \\
& \\
\displaystyle   -\frac{s_i s_j}{\sqrt{v_i v_j}{\cal S} (1-s_i^2)(1-s_j^2)} &  \text{ if $i \neq j$}.
\end{cases}
\end{equation}

To obtain the asymptotic value of $I_F$ given in~\eqref{E:asymean}
from Eqs.~\eqref{E:mean1} and~\eqref{E:fisher2}, first note that
${\mathcal S} = O(N)$.    Therefore, for large $N$,
$$
\sum_{i,j}^N R_i R_j \left[ \sqrt{v_i v_j} Q^{-1}_{i,j} \right]^2 \sim
\sum_{i}^N \left[ R_i v_i Q^{-1}_{i,i} \right]^2.
$$
Using this observation together with the asymptotic value of $Q^{-1}_{i,i}$
given in~\eqref{E:inverse_asymptot}, the first, and last two sums on the right
hand side of~\eqref{E:fisher2} behave asymptotically as
$$
\sum_{i=1}^N  (Z_i)^2 \left[ 1 + Q^{-1}_{i,i}v_i  (1 - s_i^2)\right]  +
2 \sum_{i,j}^N R_i R_j \left[ \sqrt{v_i v_j} Q^{-1}_{i,j} \right]^2 -
 4 \sum_{i=1}^N Z_i R_i v_i Q^{-1}_{i,i} \sim 2 \sum_{i=1}^N \left(Z_i - \frac{R_i}{1-s_i^2}\right)^2.
$$

By a slight  abuse of notation define the weighted average of the entries in the 
vector ${\bf a}$ over the population as
$$
\frac1N \sum_{i}^N  \frac{a_i^2}{ 1-s_i^2} = \left\langle  {\bf \frac{a^2}{1-s^2} } \right\rangle,
$$
and let
\begin{equation} \label{E:Dterm}
 D({\bf a, s}) \defn N \left[
{\bf \left\langle  \frac{a^2}{1-s^2} \right\rangle -  \left\langle  \frac{a s}{1-s^2} \right\rangle^2 \slash \left\langle {\bf \frac{s^2}{1-s^2}} \right\rangle }
\right].
\end{equation}

Then the  observations above can be combined with
\begin{equation} \label{E:defnD}
 \begin{split}
\sum_{i,j}^N a_i a_j \sqrt{v_i v_j} Q^{-1}_{i,j} &
\approx  \left[ \left( \sum_{i}^N  \frac{a_i^2}{ 1-s_i^2} \right) \left( \sum_{j}^N  \frac{s_j^2}{ 1-s_j^2} \right) -
 \left( \sum_{i}^N  \frac{a_i s_i } {1-s_i^2} \right)^2 \right] \slash \left( \sum_{j}^N  \frac{s_j^2}{ 1-s_j^2} \right) \\
& = N \left[
{\bf \left\langle  \frac{a^2}{1-s^2} \right\rangle
       \left\langle  \frac{s^2}{1-s^2} \right\rangle  -  \left\langle  \frac{a s}{1-s^2} \right\rangle^2 }
\right]
\slash \langle {\bf \frac{s^2}{1-s^2}} \rangle \\
&\defn D({\bf a, s}).
\end{split}
\end{equation}
applied to the term $\IFF$ and
the second sum on the right hand side of~\eqref{E:fisher2}, gives
\begin{equation}
\label{E:asymean}
\IFF(x) \sim  D(\frac{{\bf f'}}{\sqrt{\bf v}}, {\bf s}), \qquad 
\text{and}
\qquad 
\IFcov(x) \sim 2 \sum_{i}^N \left( \frac{v_i'}{2 v_i} - \frac{s_i' s_i}{1-s_i^2}  \right)^2 + D({\bf G s}, {\bf s}),
\end{equation}
where $G_i = \frac{d}{dx} \ln (s_i \sqrt{v_i}) = s'_i/s_i + \frac12 v'_i/v_i$.  
As before, $\IFF$  corresponds to the linear Fisher information. 

The Cauchy inequality can be applied directly
to show that
$$
{\bf \left\langle  \frac{a^2}{1-s^2} \right\rangle
       \left\langle  \frac{s^2}{1-s^2} \right\rangle  -  \left\langle  \frac{a r}{1-s^2} \right\rangle^2 } \geq 0,
$$
so that  $D(\cdot, {\bf s})$  is always positive.

\begin{figure} 
\begin{center}
\includegraphics*[0,0][272,220]{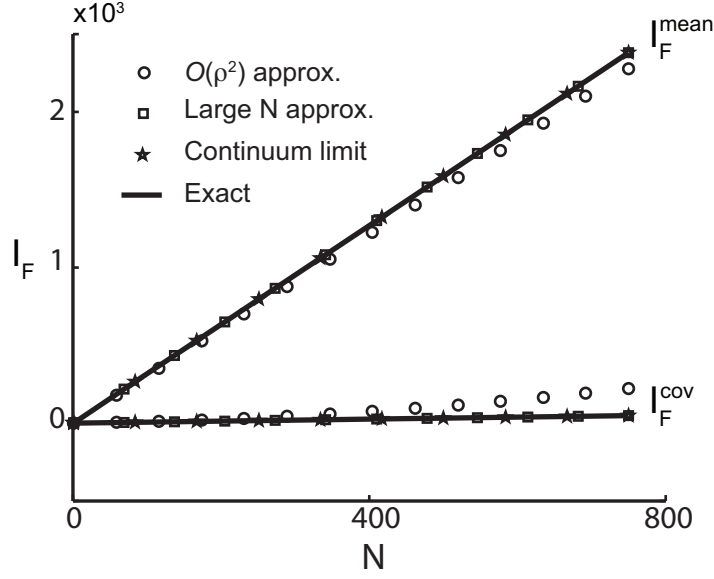}
\end{center}
\caption{Values of $\IFF$ and $\IFcov$ from:  {\it i}) approximations for small $\rho$, valid for intermediate population sizes $N$, given by Eq.~\eqref{E:smallrhopop}  {\it ii}) the ``exact" value obtained by numerically inverting the correlation matrix ${\bf Q}$, and using Eqs.~(\ref{E:IFFgeneral}--\ref{E:fullfisher}), {\it iii}) the large $N$ approximation given by Eq.~\eqref{E:asymean}, and {\it iv}) the continuum limit given by Eq.~(\ref{E:asymp}--\ref{E:continuumlimit}).     Here, $f(\phi)=5 + 45 a(\phi)$ with $a(\phi) = 1/2 (1 + \cos(\phi))$, and $v(\phi)=f(\phi)$ (as for Poisson variability).  Additionally, $s(\phi) = 0.2 + 0.5 a(\phi) \;.$  Other parameter choices give similar results (not shown).  
} \label{F:check}
\end{figure}

Figure~\ref{F:check} shows that the approximations, together with the continuum limit expressions found in the main text, are valid to high accuracy over broad ranges of $N.$

\section{Impact of pure correlation stimulus dependence on $\IFcov$.}  

\label{s.ifcov}

We show that the impact of stimulus dependence of correlations on $\IFcov$
is relatively small compared to the impact on $\IFF$ in the situation
discussed in Section~\ref{S:arbitrary}.
  By  invoking the symmetry
 of the tuning curves again
 \begin{equation} \label{E:DSr}
 D(G s, s) = D(s'(\theta) + \frac{v'(\theta)}{v(\theta)} s(\theta), s(\theta))
 \sim  \frac{N}{2\pi} \int_0^{2 \pi} \left(s'(\phi) + \frac{v'(\phi)}{v(\phi)} s(\phi)\right)^2\frac{1}{1-s^2(\phi)} d\phi,
\end{equation} 
where $s'(\theta)$ is typically much smaller than $s(\theta)v'(\theta)/v(\theta)$.  The term $D( G s,  s)$ appearing in $\IFcov$ is therefore of second order in $s(\theta)$ 
and hence negligible compared to $\IFF$.  For typical parameters, the difference is greater than an order of magnitude.

The last term in the Fisher information comes from the sum in
 $\IFcov$ given by Eq.~\eqref{E:asymean}.  In the continuum limit this term
is approximately
$$
\frac{N}{2 \pi} \int_0^{2\pi} \left( \frac{v(\phi)}{2v'(\phi)} -\frac{s'(\phi)s(\phi)}{1 - s^2(\phi)}\right)^2 d\phi
= \frac{N}{4\pi}\int_0^{2\pi} \left( \frac{d}{d\phi} \left[\log (v(\phi) (1 - s^2(\phi))\right] \right)^2 d\phi.
$$
For the type of stimulus dependence that we assume  $\frac{v(\phi)}{2v'(\phi)}$ and $-\frac{s'(\phi)s(\phi)}{1 - s^2(\phi)}$ have opposite
signs.  For small correlations, the first term will dominate and stimulus dependence of correlations will
decrease this entry in $\IFcov$.  When correlations are not perfect (near 1) the term $\IFV$ is typically much smaller than $\IFF$.

\section{Details of the numerical implementations}
\label{A:constantcorrelations}

Numerical values of Fisher Information in Figs.~\ref{F:nullfull1} and \ref{F:nullfull2} were found by directly inverting the correlation matrices $Q$ and performing the required matrix multiplications in MATLAB.  The authors are happy to provide these codes upon request.  

The procedure is as follows:  We first fix the average value of correlations, $\langle \rho_{ij} \rangle$, among all neurons in the population (the value $\langle \rho_{ij} \rangle=0.1$ was used for all figures in this paper).  Next, we define correlation matrices  consistent with this value of $\langle \rho_{ij} \rangle$, for two cases,  {{\it Stimulus Dependent (SD)}} and {{\it Stimulus Independent (SI)}} (see main text). We first define $Q_{i,j}$ via Eqn.~\eqref{E:crosscor}, assuming that the $\rho_{i,j} (\theta)$ are given by~\eqref{E:rij}.  Here, for Figs.~\ref{F:nullfull1} and \ref{F:nullfull2}, we used  $s(\theta)$ =$ k_\rho + b_\rho a^2(\theta)$, where $a(\theta) = 1/2 (1 + \cos(\theta))$ and $ k_\rho$ and $b_\rho$ are constants chosen as follows:  (i) the average correlation $\langle \rho_{ij} \rangle=0.1$, and (ii) the ratio of largest so smallest pairwise correlations, $(k_\rho + b_\rho)^2/b_\rho^2$, should be $R=10$ for the SD case and $R=0$ (i.e., $b_\rho=0$) for the SI case.

To study the affects of heterogeneity, as a final step we jitter the tuning curves for $s$ and $v$ by $\pm 20 \%$.

\bibliographystyle{plain}

\bibliography{info2}






\end{document}